\documentclass{article}
\usepackage[latin9]{inputenc}
\usepackage{geometry}
\usepackage{amsmath}
\usepackage{amssymb}
\usepackage[unicode=true,pdfusetitle,
 bookmarks=true]
 {hyperref}

\makeatletter

\providecommand{\tabularnewline}{\\}
\usepackage{float}
\usepackage{graphicx}
\usepackage{grffile}
\usepackage{breakurl}
\usepackage{caption}

\providecommand{\tabularnewline}{\\}
\usepackage{fullpage}
\usepackage[justification=centering]{caption}
\usepackage[all]{hypcap}
\usepackage{microtype}% % % % % % % % % % % % % % % % % % % % % BEGIN DOCUMENT

\makeatother

\begin{document}

\title{Higher Shell Configuration Mixing for Magnetic Moments}

\author{Xiaofei Yu$^{1}$, Larry Zamick$^{1}$ and Shadow Robinson$^{2}$\\
 \textit{1.Department of Physics and Astronomy, Rutgers University,
Piscataway, New Jersey 08854}\\
 \textit{2.Department of Physics, Millsaps College, Jackson, Mississippi
39210} }

\date{\today}
\maketitle
\begin{abstract}
In many works on nuclear magnetic moments, shell model calculations
must ignore certain configurations since the result in a model space
is too large. Motivated by this we here construct a simple model which
will enable us to evaluate the effects of high lying configurations.
We start with 2 neutrons in the $g_{7/2}$ single shell and then admix
systematically other configurations: $h_{11/2}h_{11/2}$ , $h_{11/2}h_{9/2}$
and $h_{9/2}h_{9/2}$ and examine how configurations affect the magnetic
moments ($g$ factors) of $J=2^{+}$, $4^{+}$ and $6^{+}$ states.
We have a simple formula which explains the qualitative behaviour
of the efffects of higher shells.
\end{abstract}

\section*{Introduction }

The nuclear magnetic moment of a free neutron in units of $\mu_{N}$
is $-1.913$ ($+2.793$ for a proton). The magnetic moment operator
is $g_{s}S+g_{l}L$. Allowing for quenching we have for a neutron
$g_{s}=-3.826x$ where for $x=1$ if there is no quenching. Also for
a neutron we have $g_{l}=0$, while for a proton $g_{l}=1$. However
often the values $g_{l}=-0.2Z/A$ is used for the neutron and $1+0.2N/A$
for the proton.

The Schmidt values for a single $j$ shell : 
\begin{equation}
g=\frac{1}{j}(g_{l}l+\frac{g_{s}}{2})\qquad\qquad\qquad\text{if}\qquad j=l+\frac{1}{2}
\end{equation}

\begin{equation}
g=\frac{1}{j+1}(g_{l}(l+1)-\frac{g_{s}}{2})\qquad\text{if}\qquad j=l-\frac{1}{2}
\end{equation}

For 2 nucleons in different shells $j_{1}$ and $j_{2}$ the g factors
depend of the total angular momentum $J=j_{1}+j_{2}$. 
\begin{equation}
g=\frac{1}{2}(g_{1}+g_{2})+(g_{1}-g_{2})\frac{j_{1}(j_{1}+1)-j_{2}(j_{2}+1)}{2J(J+1)}
\end{equation}

Where $g_{1}$ and $g_{2}$ are $g$ factors of nucleon 1 and nucleon
2. We will use equation (1) and (2) to calculate the $g$ factors
of single nucleon state and equation (3) for $g$ factors of two mixed
nucleon state.

Often when large scale shell model calculations of nuclear magnetic
moments are performed, one is frustrated by the fact that one cannot
include configurations that might be important, because the model
space becomes too large to handle. One example in which some of the
current authors were involved is the work of Kumartzki et al\cite{Kumbartzki}.

In this work we will consider a very simple system of 2 neutrons in
various shells. We start with the configuration $[g_{7/2}g_{7/2}]^{J}$
with $J=2^{+}$, $4^{+}$ and $6^{+}$. The $g$ factors for all 3
J's are the same and are equal to the single particle g factor in
the $g_{7/2}$ shell. The free value is 0.42511. We will see how this
number changes when we use a more elaborate wave function: 
\begin{equation}
\Psi^{J}=a{[}g_{7/2}g_{7/2}{]}^{J}+b{[}h_{11/2}h_{11/2}{]}^{J}+c\frac{\sqrt{2}}{2}([h_{11/2}(1)h_{9/2}(2){]}^{J}-{[}h_{11/2}(2)h_{9/2}(1){]}^{J})+d{[}h_{9/2}h_{9/2}{]}^{J}
\end{equation}

The four terms in the wave function associated with $a$, $b$, $c$,
$d$ are chosen to be the basis states and $|a|^{2}$ is the probability
finding the system in the state of $[g_{7/2}g_{7/2}]^{J}$, $|b|^{2}$
is the probability finding the system in the state of $[h_{11/2}h_{11/2}]^{J}$
etc.. (Note, since the states on interest have positive parity, there
are no $[g_{7/2},h_{j}]^{J}$ terms). The $g$ factors of the basis
states are respectively $+0.42511$ (from equation (2)), $-0.34782$
(from equation (1)), $-0.63767/J(J+1)$ (from equation (3)) and +0.34782
(from equation (2)). From equation (1) and (2) ($g_{l}=0$ for neutron),
we see that the $g$ factor in single state $h_{9/2}$ is equal but
opposite to that of $h_{11/2}$. Equation (3) implies that $g$ factor
for the mixed state of $h_{11/2}$ and $h_{9/2}$ is $J$ dependant.

\section*{Calculations}

We will obtain the wave functions and $g$ factors using a surface
delta interaction. This was used previously by us for $g$ factor
calculations in $^{86}\mathrm{Kr}$ and $^{112}\mathrm{Sn}$ \cite{Zamick,Yu}.

The $\mathrm{T}=1$ matrix element of the SDI interaction can be written
as following \cite{Green,Arvieu,Talmi}: 
\begin{equation}
<j_{1}j_{2}|SDI|j_{3}j_{4}>=C_{0}f(j_{1},j_{2})f(j_{3},j_{4})
\end{equation}
where $f(j_{1},j_{2})=(-1)^{j_{2}+\frac{1}{2}}\sqrt{\frac{(2j_{1}+1)(2j_{2}+1)}{(2J+1)(1+\delta_{j_{1}j_{2}})}}\cdot\Big<j_{1}j_{2}\text{\space}\frac{1}{2}\left(\frac{\text{-}1}{2}\right)\Big|J0\Big>$

Note that the expression is separable. In previous works \cite{Zamick,Yu}
we chose $C_{0}$ to be $-0.200M\mathrm{e}V$. We begin in Table I
by having the single particle splitting for $g_{7/2}-h_{11/2}$ and
$g_{7/2}-h_{9/2}$ be identical and show results for a splitting $E$
with values of $E=+0.5M\mathrm{e}V$ and $E=+1.0M\mathrm{e}V$. It
is really the ratio $E/C_{0}$ that is the relevant parameter.

\begin{center}
\captionof{table}{$g$ factor as a function of $E$ and $J$ with
different configurations} %
\begin{tabular}{lccc}
 &  &  & \tabularnewline
\hline 
Configuration  & $J$  & $g(E=0.5)$  & $g(E=1)$\tabularnewline
\hline 
 &  &  & \tabularnewline
Case 1: $c=d=0$  & 2  & 0.2870  & 0.3805\tabularnewline
 & 4  & 0.3767  & 0.4128\tabularnewline
 & 6  & 0.4085  & 0.4212\tabularnewline
 &  &  & \tabularnewline
Case 2: $c=0$  & 2  & 0.2497  & 0.3599\tabularnewline
 & 4  & 0.3536  & 0.4085\tabularnewline
 & 6  & 0.4017  & 0.4202\tabularnewline
 &  &  & \tabularnewline
Case 3: $a,b,c,d\not=0$  & 2  & 0.2015  & 0.3394\tabularnewline
 & 4  & 0.3100  & 0.3988\tabularnewline
 & 6  & 0.3656  & 0.4146\tabularnewline
 &  &  & \tabularnewline
Initial Case: $[g_{7/2}g_{7/2}]$  & All $J$  & 0.4251  & 0.4251\tabularnewline
\hline 
\end{tabular}
\par\end{center}

In case 1, we only include $[h_{11/2}h_{11/2}]^{J}$ and $[g_{7/2}g_{7/2}]^{J}$,
since $c$ and $d$ are set to be 0. For Case 2, we will also include
$[h_{9/2}h_{9/2}]^{J}$, that is, only $c$ is set to zero. Lastly
in Case 3, we include all 4 configurations, that is $a$, $b$, $c$,
and $d$ are not 0. Let us first focus on the $g$ factors of $J=2^{+}$
state. The $g$ factor of Initial Case is $0.4251$, and it reduces
to $0.2870$ when we include $[h_{11/2}h_{11/2}]^{J}$,($c=d=0$)
i.e. a 32.5\% reduction. This is due to the fact that the $[h_{11/2}h_{11/2}]^{J}$
state has a $g$ factor that is the opposite sign of that of $[g_{7/2}g_{7/2}]^{J}$.
In general a $g$ factor of neutron in $j=l+1/2$ state has the opposite
sign from that of $j=l-1/2$ state.

We then set $c=0$. Here is where the counterintuitive behavior appears.
The $g$ factor of the $h_{9/2}$ neutron is equal and opposite to
that of a $h_{11/2}$ neutron. This would suggest that $h_{9/2}$
would undo the damage done by $h_{11/2}$ leading to a $g$ factor
closer to that of a pure $[g_{7/2}g_{7/2}]$ configuration. But the
opposite is true. The value in this case is 0.2497 as compared with
0.2870.

To understand what is happening, we show the detailed wave functions
in Table II for the case $E=+0.5$. The important point is that the
amplitude $a$ is much smaller in Case 2. Introducing the configuration
$[h_{9/2}h_{9/2}]^{J}$ causes the $[g_{7/2}g_{7/2}]^{J}$ probability
to be depleted. This depletion causes the overall g factor to be come
smaller. However the $[h_{11/2}h_{11/2}]^{J}$ component does not
get depleted. In fact it is slightly enhanced. Clearly these effects
are beyond first order perturbation theory.

We note that the $g$ factors, which were $J$ independent for the
configuration $[g_{7/2}g_{7/2}]^{J}$ are now all different, with
$g(J=2)$ is the smallest, $g(J=4)$ is in the middle, and $g(J=6)$
is the largest.

We note that often experiments are performed for nuclei which are
difficult to handle theoretically within the shell model framework.
We view this work as exploratory. By starting with the simplest configurations,
we can include configurations that are not possible for large scale
shell model calculations. The way the high lying configurations affect
the physical properties (in this case $g$ factors) in this simple
case gives us a clue of what is missing in the more complex situations.

\begin{center}
\captionsetup{width=0.5\linewidth} \captionof{table}{Wave
function coefficients of Eq (4) with E=+0.5 for Case 1, Case 2, and
Case 3} %
\begin{tabular}{llll}
\hline 
 & Case 1  & Case 2  & Case 3\tabularnewline
\hline 
$a$  & -0.9063  & -0.7857  & -0.7711\tabularnewline
$b$  & 0.4227  & 0.4570  & 0.4574\tabularnewline
$c$  & 0  & 0  & -0.1418\tabularnewline
$d$  & 0  & 0.4159  & 0.4174\tabularnewline
\hline 
\end{tabular}\\

\par\end{center}

In Table III we show the values of $g(J=2)$ for various values of
$E$. The behavior is a bit complex. Up to $E=+0.2$ the values in
Case 1 ($[h_{11/2}h_{11/2}]$ included) are smaller than those in
Case 2 (both $[h_{11/2}h_{11/2}]$ and $[h_{9/2}h_{9/2}]$ included).
But small $E$ corresponds to strong coupling. For $E=+0.3$ and beyond
we get a reversal with Case 2 yielding $g$ factors smaller than that
of Case 1. But higher $E$ means weaker coupling, so one might have
expected the opposite behavior. In Table II there is a considerable
depletion of the $[g_{7/2}g_{7/2}]$ configuration when one goes from
Case 1 ($|a|^{2}=82\%$) to Case 2 (only 62\%).

In Table IV we show the results with all the configurations present.
We do it for Case 3 as before, where the $h_{11/2}$ and $h_{9/2}$
are degenerate at and energy $E$ above $g_{7/2}$ and a new Case
4 where $h_{11/2}$ is at an energy $E$ above $g_{7/2}$ and $h_{9/2}$
is raised above $h_{11/2}$ and at an energy $E+0.25$.

\qquad{}

\begin{center}
\begin{minipage}[c]{0.4\linewidth}%
\begin{center}
\captionsetup{width=0.8\linewidth} \captionof{table}{$g(2^{+}$)
for increasing $E$ in Case 1 and Case 2} 
\par\end{center}

\begin{center}
\begin{tabular}{lll}
\hline 
$E$  & Case 1  & Case2\tabularnewline
\hline 
0  & -0.04375  & 0.08689\tabularnewline
0.1  & 0.03601  & 0.11369\tabularnewline
0.2  & 0.11610  & 0.14496\tabularnewline
0.3  & 0.18702  & 0.17959\tabularnewline
0.4  & 0.24403  & 0.21554\tabularnewline
0.5  & 0.28700  & 0.24972\tabularnewline
1  & 0.38047  & 0.35988\tabularnewline
2  & 0.41362  & 0.40971\tabularnewline
3  & 0.42006  & 0.41874\tabularnewline
4  & 0.42230  & 0.42168\tabularnewline
5  & 0.42332  & 0.42298\tabularnewline
INF.  & 0.42511  & 0.42511\tabularnewline
\hline 
\end{tabular}
\par\end{center}%
\end{minipage}%
\begin{minipage}[c]{0.4\linewidth}%
\begin{center}
\captionsetup{width=1.1\linewidth} \captionof{table}{$g(2^{+})$
for increasing $E$ in Case 3 and Case 4 where $E(h_{9/2})=E(h_{11/2})+0.25$.} 
\par\end{center}

\begin{center}
\begin{tabular}{lll}
\hline 
$E$  & Case 3  & Case 4\tabularnewline
\hline 
0  & 0.01276  & -0.14372\tabularnewline
0.1  & 0.04309  & -0.07784\tabularnewline
0.2  & 0.07871  & 0.00700\tabularnewline
0.3  & 0.11854  & 0.06381\tabularnewline
0.4  & 0.16402  & 0.12953\tabularnewline
0.5  & 0.20152  & 0.18660\tabularnewline
1  & 0.33984  & 0.34157\tabularnewline
2  & 0.40511  & 0.40514\tabularnewline
3  & 0.41690  & 0.41683\tabularnewline
4  & 0.42071  & 0.43064\tabularnewline
5  & 0.42238  & 0.42235\tabularnewline
INF.  & 0.42511  & 0.42511\tabularnewline
\hline 
\end{tabular}
\par\end{center}%
\end{minipage}
\par\end{center}

We would expect Case 4 to give smaller values than Case 3, because
we have raised the energy of the $h_{9/2}$ energy relative to $h_{11/2}$.
This is indeed true, up to $E=0.5$ but for $E=1$ and $E=2$ there
is a reversal. But for $E=4$ and $E=5$ we are back to the ordering
for $E$ equal or less than 0.5. It is not surprising that as $E$
getting very large, $g(2^{+})$ approaches to the singe particle value
of $g_{9/2}$, which equals 0.42511.

\section*{Explanation}

To attempt an explanation of the results we compare the results for
Case 1 (We will denote $G$ as $g$ factor to avoid confusion between
$g$ of $g$ factor and $g$ of $g_{7/2}$ state.

In Case 1 the wave function is 
\begin{equation}
\Psi_{1}^{J}=a_{1}[g_{7/2}g_{7/2}]^{J}+b_{1}[h_{11/2}h_{11/2}]^{J}
\end{equation}

The $g$ factor in Case 1 is 
\begin{equation}
G_{1}(J)=a_{1}^{2}G(g{7/2})+b_{1}^{2}G(h_{11/2})
\end{equation}

We then use the fact that $a_{1}^{2}+b_{1}^{2}=1$ to rewrite equation
(7) as 
\begin{equation}
G_{1}(J)=G(g_{7/2})+b_{1}^{2}(G(h_{11/2})-G(g_{7/2}))
\end{equation}

The wave function for Case 2 is 
\begin{equation}
\Psi_{2}^{J}=a_{2}[g_{7/2}g_{7/2}]+b_{2}[h_{11/2}h_{11/2}]+d_{2}[h_{9/2}h_{9/2}]
\end{equation}

The $g$ factor is therefore given by following 
\begin{equation}
G_{2}(J)=a_{2}^{2}G(g_{7/2})+b_{2}^{2}G(h_{11/2})+d_{2}^{2}G(h_{9/2})
\end{equation}

Since $a_{2}^{2}+b_{2}^{2}+d_{2}^{2}=1$, and $G(h_{11/2})=-G(h_{9/2})$
then we rewrite equation (10) as 
\begin{equation}
G_{2}(J)=(1-b_{2}^{2}-d_{2}^{2})G(g_{7/2})+(b_{2}^{2}-d_{2}^{2})G(h_{11/2})
\end{equation}

Then using equation (8) to subtract equation (11), we have

\begin{equation}
G_{1}(J)-G_{2}(J)=(b_{2}^{2}+d_{2}^{2}-b_{1}^{2})G(g_{7/2})+(b_{1}^{2}-b_{2}^{2}+d_{2}^{2})G(h_{11/2})
\end{equation}

In perturbation theory we have $b_{1}=b_{2}$. Using this we find
\begin{equation}
G_{1}(J)-G_{2}(J)=d_{2}^{2}(G(g_{7/2})+G(h_{11/2}))
\end{equation}
Since $G(g_{9/2})=1.913/4.5$ and $G(g_{11/2})=-1.913/5.5$, then
$G_{1}(J)-G_{2}(J)=+0.07729d_{2}^{2}$

We see that Case 1 gives a larger value ( closer to the pure g$_{7/2}$$^{2}$
case) than Case 2 for all $d_{2}^{2}$ in this approximation. For
the weak coupling or strong coupling, we get the same qualitative
behavior i.e.introducing $h_{9/2}$ to the existing $h_{11/2}$ admixture
does not cancel out the effect of the $h_{11/2}$ admixture but rather
enhances it.In paer this is due to the fact that the g factor for
h$_{9/2}$,although positive, is smaller than the g factor for g$_{7/2}$.

We can simulate perturbation theory by simply setting $a=1$. Using
$E=0.5$ as an example. In the \char`\"{}exact\char`\"{} calculation,
the values of G(J=2) in the 3 cases were respectively 0.2870, 0.2497,
0.2015, now with $a=1$ we obtain 0.3629, 0.4123, 0.3738. Whereas
in the case where $a$ is less than one we go from large to smaller
to smallest, in the perturbation calculation ($a=1$) we go from small
to large to smaller. There is a qualitative difference.

We repeat that it is of value to study simple systems. In this case
we examine the 2 neutron system, because it enables us to include
configurations that are at present not possible for more complicated
systems. We can get some idea of what the missing elements of these
more complex system do.

\section*{Appendix: Expressions for the $G$ factors}

Again, we use the symbol $G$ for the $g$ factors so as to distinguish
them form the $g$ in $g_{7/2}$. Note that in a single $j$ shell
of particles of one kind, all $g$ factors are the same. In particular
for this case the $g$ factor of two $g_{7/2}$ is the same as that
of one. The same is true for $h_{9/2}$ and $h_{11/2}$.

The expression for the $g$ factors is:

\begin{equation}
G(J)=a^{2}G(g_{7/2})+b^{2}G(h_{11/2})+c^{2}G(h_{11/2}h_{9/2})+d^{2}G(h_{9/2})+CrosstermA+CrosstermB
\end{equation}

We then use equation (3) to expand $G(h_{11/2}h_{9/2})$ 
\begin{equation}
G(h_{11/2}h_{9/2})=\frac{1}{2}(G(h_{11/2})+G(h_{9/2}))+(G(h_{11/2})-G(h_{9/2}))\frac{11/2\cdot13/2-9/2\cdot11/2}{2J(J+1)}
\end{equation}

\begin{equation}
CrossA=4bc\cdot U(1,4.5,J;5.5,5.5,J)\frac{<h_{11/2}||\mu||h_{9/2}>}{\sqrt{2(J(J+1)}}
\end{equation}

\begin{equation}
CrossB=-4cd\cdot U(1,5.5,J,4.5;4.5,J)\frac{<h_{9/2}||\mu||h_{11/2}>}{\sqrt{2J(J+1)}}
\end{equation}
where $U$ is a unitary Racah coefficient and at the end we have reduced
matrix elements of the magnetic moment operator $\mu=g_{L}L+g_{S}\sigma/2$.

The numerical values for bare operators which can be obtained from
equation (1), (2), and (3) are: $G(g_{7/2})=0.4251$, $G(h_{11/2})=-0.3478$,
$G(h_{9/2})=+0.34782$. $G(h_{11/2}h_{9/2})=-3.8260/(J(J+1))$. The
respective values for $J=2$, 4, and 6 are $-0.6377$, $-0.1913$
and $-0.0911$

The bare values are $g_{L}=0$ and $g_{S}=-3.826$. Our reduced matrix
element is one used by Lawson \cite{Lawson} : 
\begin{equation}
<\psi_{M_{B}}^{J_{B}}O_{\mu}^{\lambda}\psi_{M_{A}}^{J_{A}}>=<J_{A}\lambda M_{A}\mu|J_{B}M_{B}><J_{B}||O^{\lambda}||J_{A}>
\end{equation}

The reduced matrix elements are 

\begin{center}
$<h_{11/2}||\sigma||h_{9/2}>=-\sqrt{20/11}$
\par\end{center}

\begin{center}
$<h_{9/2}||\sigma||h_{11/2}>=\sqrt{24/11}$
\par\end{center}

\begin{center}
$<h_{11/2}||L||h_{9/2}>=\sqrt{5/11}$
\par\end{center}

\begin{center}
$<h_{9/2}||L||h_{11/2}>=-\sqrt{6/11}$ 
\par\end{center}

The one used by Bohr and Mottelson is $<\psi_{M_{B}}^{J_{B}}O_{\mu}^{\lambda}\psi_{M_{A}}^{J_{A}}>=<J_{A}\lambda M_{A}\mu|J_{B}M_{B}>\frac{<J_{B}||O^{\lambda}||J_{A}>}{\sqrt{2J_{B}+1}}$
\cite{Bohr}.

\section*{Acknowledgement}

X.Y. acknowledges the support from Aresty program in Rutgers University-New
Brunswick for 2015-2016 academic year. Xiaofei Yu thanks Xiangyi Shan
for useful discussion on the format of this paper and the code programming.

\newpage{}

\end{document}